\documentclass[twocolumn,prb]{revtex4-2}
\usepackage{amsfonts}
\usepackage[T1]{fontenc}
\usepackage{amsmath,amsbsy,amssymb,graphicx}
\usepackage{times}

\begin{document}

\title{Nonlinear dynamical topological phases in Cooper-pair box array}
\author{Motohiko Ezawa}
\affiliation{Department of Applied Physics, University of Tokyo, Hongo 7-3-1, 113-8656,
Japan}

\begin{abstract}
The topological property of a system is a static property in general. For
instance, the topological edge state is observed by measuring the local
density of states. In this work we propose a system whose topological
property is only revealed by dynamics. As a concrete example, we explore a
nonlinear dynamical topological phase transition revealed by a quench
dynamics of a Cooper-pair box array connected with capacitors. It is
described by coupled nonlinear differential equations due to the Josephson
effect. It is trivial as far as the static system is concerned. However, the
wave propagation induced by the quench dynamics demonstrates a rich
topological phase diagram in terms of the strength of the input.
\end{abstract}

\maketitle

Topological physics is one of the most extensively studied fields in
condensed-matter physics\cite{Hasan,Qi}. The notion of topology is also
applicable to artificial topological systems such as photonic\cite%
{KhaniPhoto,Hafe2,Hafezi,WuHu,TopoPhoto,Ozawa}, acoustic\cite%
{Prodan,TopoAco,Berto,Xiao,He}, mechanical\cite%
{Lubensky,Chen,Nash,Paul,Sus,Sss,Huber,Mee} and electric circuit\cite%
{TECNature,ComPhys,Hel,Lu,YLi,EzawaTEC} systems. New feature of these
artificial topological systems are that nonlinearity is naturally introduced
as in photonics\cite{Ley,Zhou,MacZ,HadadB,Smi,Tulo,Kruk,NLPhoto,Kirch},
mechanics\cite{Snee,PWLo,MechaRot} and electric circuits\cite%
{Hadad,Sone,TopoToda}. In particular, electric circuits simulate almost all
topological phases. Active topological electric circuits\cite{ATC} is
realized by using nonlinear Chua's diode circuit. Topological Toda lattice
is realized by using variable capacitance diodes\cite{TopoToda}.

\begin{figure}[t]
\centerline{\includegraphics[width=0.48\textwidth]{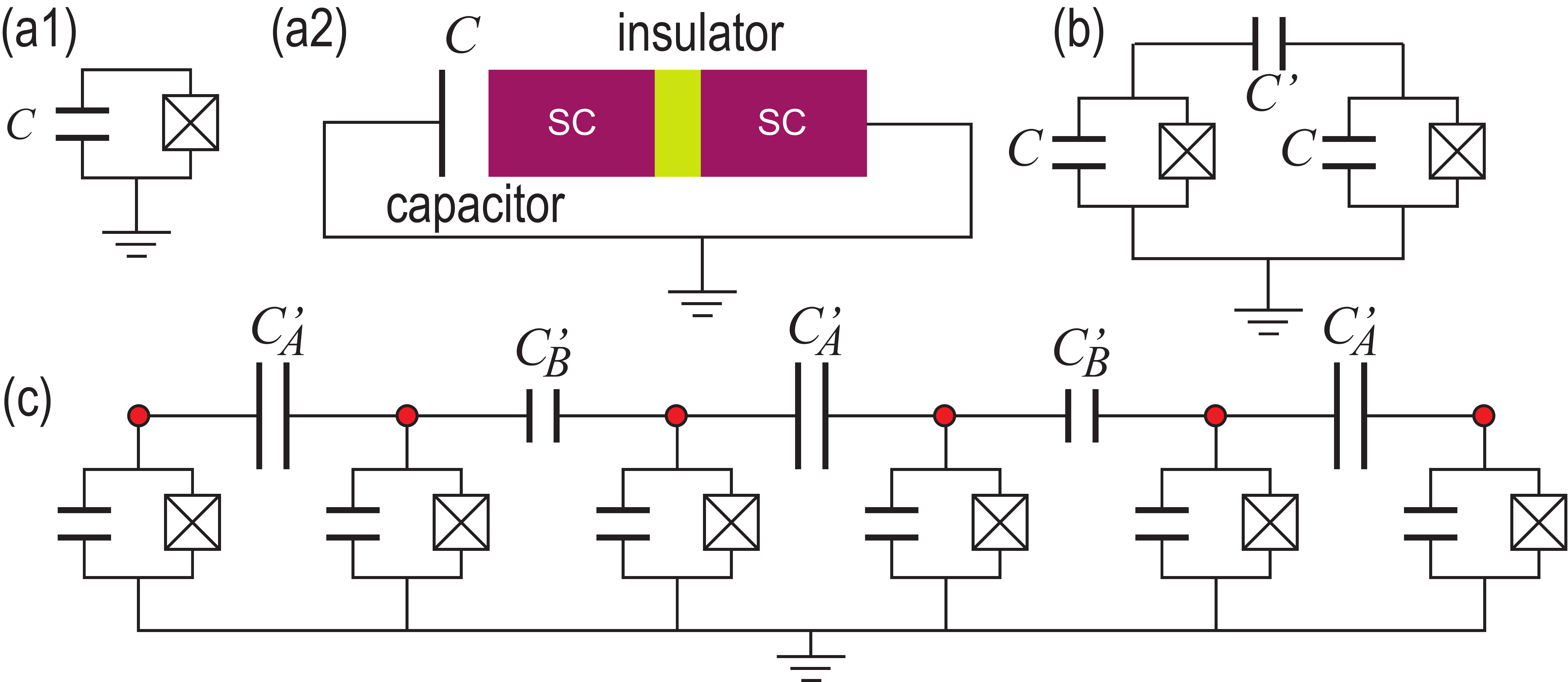}}
\caption{Illustration of (a) a Cooper-pair box, (b) double Cooper-pair boxes
connected via a capacitor and (c) a dimerized Cooper-pair box array.}
\label{FigCPBox}
\end{figure}

Recently, the development of superconducting qubits is very rapid as a
method for quantum computation\cite{Nakamura,Google}. The transmon qubit is
a successful example of a superconducting qubit\cite{Koch,Schr} based on a
Cooper-pair box (CPB)\cite{Nakamura,Bouch,MakRev} made of the Josephson
junction and a capacitor, as shown in Fig.\ref{FigCPBox}(a).

In this work we explore a nonlinear topological phase transition in a CPB
array connected with capacitors as in Fig.\ref{FigCPBox}(d), where
alternating capacitances are arranged for the array to acquire a structure
akin to that of the one-dimensional Su-Schrieffer-Heeger (SSH) chain. The
SSH model is the well-known topological system, where the topological phase
is signaled by the emergence of zero-energy topological edge states. They
are detected by the measurement of the local density of states. However, the
CPB array reveals an unfamiliar feature in topological physics, where the
system is trivial as far as the static system is concerned.

To reveal the topological structure, we evoke a dynamical property. We
analyze a quench dynamics by giving a flux to the left-end CPB and studying
its time evolution. We have found four types of propagations. (i) There are
standing waves mainly at the left-end CPB and weakly at a few adjacent
odd-number CPBs. Additionally, there are propagating waves into the bulk.
(ii) There are only propagating waves spreading into other CPBs. (iii) The
standing wave is trapped strictly at the left-end CPB. (iv) The coupled
standing waves are trapped to double CPBs at the left-end. These four modes
correspond to the topological phase, the trivial phase, the trapped phase
and the dimer phase. We determine the phase diagram with the aid of a phase
indicator defined by the saturated amplitude. The present model presents a
system whose topological property is only revealed by a dynamical method.

\textbf{Model:} A CPB is an electric circuit made of a Josephson junction
and a capacitor with capacitance $C$, as illustrated in Fig.\ref{FigCPBox}(a). 
We investigate a CPB array connected by capacitors shown in Fig.\ref{FigCPBox}(c). The Lagrangian is given by%
\begin{align}
\mathcal{L(}\Phi ,\dot{\Phi})=& \sum_{n}\left[ \frac{C_{n}}{2}\dot{\Phi}_{n}^{2}
+\frac{I_{\text{c}}\Phi _{0}}{2\pi }\cos 2\pi \frac{\Phi _{n}}{\Phi
_{0}}\right]   \notag \\
& +\sum_{n}\frac{C_{n}^{\prime }}{2}(\dot{\Phi}_{n}-\dot{\Phi}_{n+1})^{2},
\label{Lag}
\end{align}%
where $I_{\text{c}}$ is the critical current of the Josephson junction, 
$\Phi _{0}\equiv h/2e$ is the unit flux, $C_{n}$ is the capacitance in the $n$-th CPB, 
and $\Phi _{n}$ is a magnetic flux across the $n$-th Josephson
junction. The first line describes the $n$-th CPB, and the second line
describes the coupling between the $n$-th and ($n+1$)-th CPBs via the
capacitance $C_{n}^{\prime }$. It is summarized as%
\begin{equation}
\mathcal{L(}\Phi ,\dot{\Phi})=\frac{1}{2}\sum_{n,m}M_{nm}\dot{\Phi}_{n}
\dot{\Phi}_{m}+\sum_{n}\frac{I_{\text{c}}\Phi _{0}}{2\pi }\cos 2\pi \frac{\Phi
_{n}}{\Phi _{0}},
\end{equation}%
where%
\begin{equation}
M_{nm}=[C_{n}+C_{n}^{\prime }+C_{n-1}^{\prime }]\delta _{nm}-[C_{n}^{\prime
}\delta _{m,n+1}+C_{m}^{\prime }\delta _{n,m+1}],
\end{equation}%
with $C_{0}^{\prime }=0$. The Euler-Lagrange equation reads%
\begin{equation}
\sum_{m}M_{nm}\ddot{\Phi}_{m}+I_{\text{c}}\sin 2\pi \frac{\Phi _{n}}{\Phi
_{0}}=0.  \label{MEq}
\end{equation}%
The canonical conjugate of the flux $\Phi _{n}=\partial \mathcal{L}/\partial 
\dot{\Phi}_{n}$ is the charge acquired in the capacitor, and given by%
\begin{equation}
Q_{n}=M_{nm}\dot{\Phi}_{m}.  \label{Charge}
\end{equation}%

\begin{figure}[t]
\centerline{\includegraphics[width=0.48\textwidth]{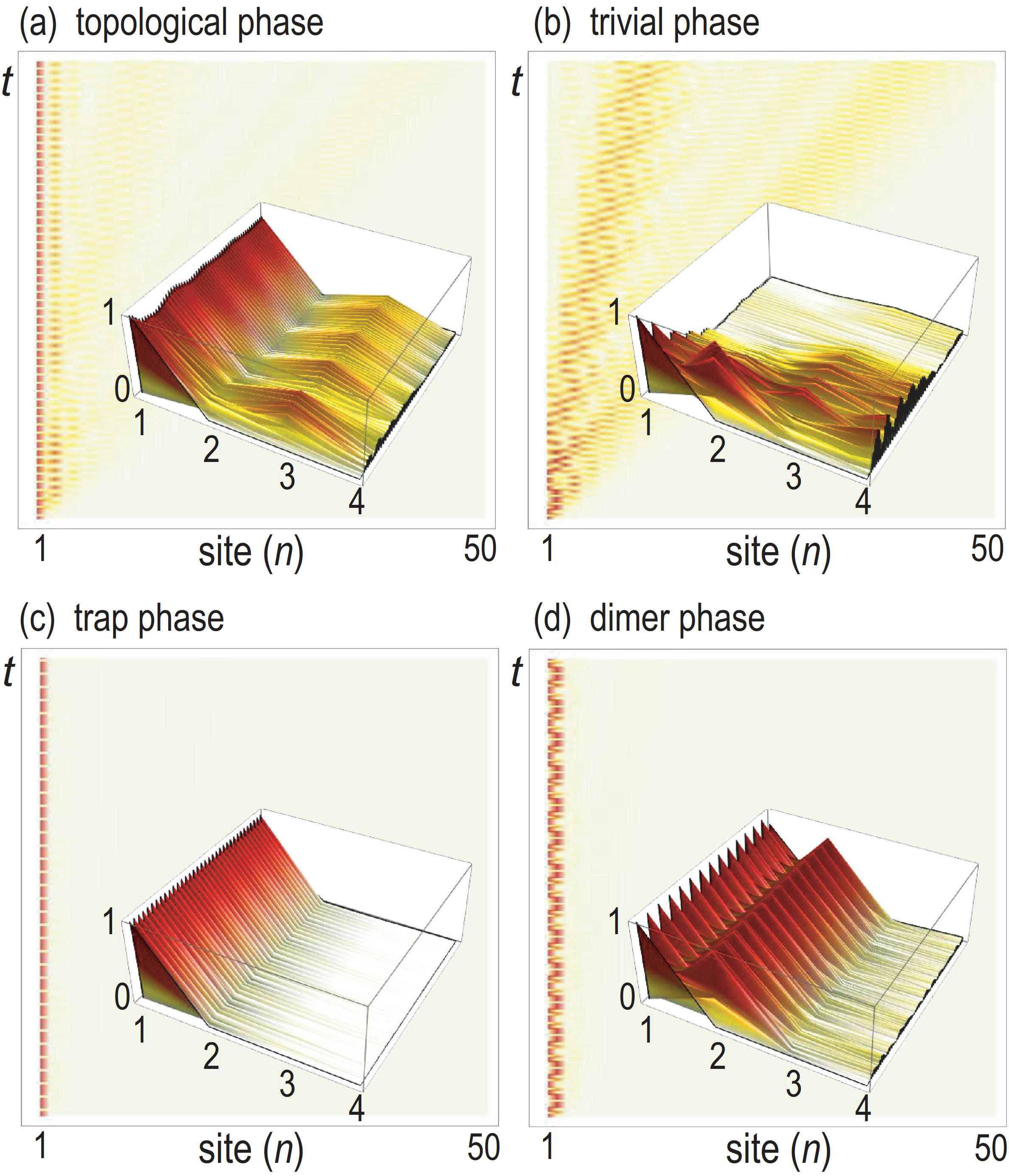}}
\caption{Time evolution of $\Phi _{n}$ in the $n$-th Cooper-pair box. (a)
The topological phase with $\protect\lambda =-0.5$ and $\protect\xi =0.25$,
where there are simple standing waves at $n=1$ and $n=3$. (b) The trivial
phase with $\protect\lambda =0.5$and $\protect\xi =0.25$, where the
oscillation propagates along the array. (c) The trap phase with $\protect\lambda =-0.8$ 
and $\protect\xi =0.8$, where the standing wave is present
only at the left-end Cooper-pair box. (d) The dimer phase with $\protect\lambda =0.9$ 
and $\protect\xi =0.5$, where coupled standing waves are
trapped to two Cooper-pair boxes at the left-end. We have used a chain with
length $100$. Insets show an enlarged dynamics with length $4$, where the
vertical axis is $\Phi _{n}$ in units of $\left( \protect\xi \Phi
_{0}/2\right) $. We have set $T=50$.}
\label{FigDynamics}
\end{figure}

\textbf{Dimerized CPB array:} To equip the system with a topological
structure, we consider a one-dimensional periodic chain, where all
capacitances within CPBs are taken equal and the capacitances connecting
CPBs are taken to be alternating, 
\begin{equation}
C_{A}^{\prime }=C_{2n+1}^{\prime }=C^{\prime }\left( 1+\lambda \right)
,\quad C_{B}^{\prime }=C_{2n}^{\prime }=C^{\prime }\left( 1-\lambda \right) ,
\end{equation}%
with the dimerization $\left\vert \lambda \right\vert \leq 1$. We write down
Eq. (\ref{MEq}) explicitly as follows,%
\begin{align}
C_{B}^{\prime }\ddot{\Phi}_{2n-2}-\left( C+2C^{\prime }\right) & \ddot{\Phi}_{2n-1}
+C_{A}^{\prime }\ddot{\Phi}_{2n}  \notag \\
& =I_{\text{c}}\sin 2\pi (\Phi _{2n-1}/\Phi _{0}), \\
C_{A}^{\prime }\ddot{\Phi}_{2n-1}-\left( C+2C^{\prime }\right) & \ddot{\Phi}_{2n}
+C_{B}^{\prime }\ddot{\Phi}_{2n+1}  \notag \\
& =I_{\text{c}}\sin 2\pi (\Phi _{2n}/\Phi _{0}).
\end{align}%
Our task is to solve these equations. The flux $\Phi _{n}$ propagates as
shown in Fig.\ref{FigDynamics}. Accordingly, the charge $Q_{n}$ also
propagates along the array with the use of the solution $\Phi _{n}$
according to the formula (\ref{Charge}).

\begin{figure}[t]
\centerline{\includegraphics[width=0.48\textwidth]{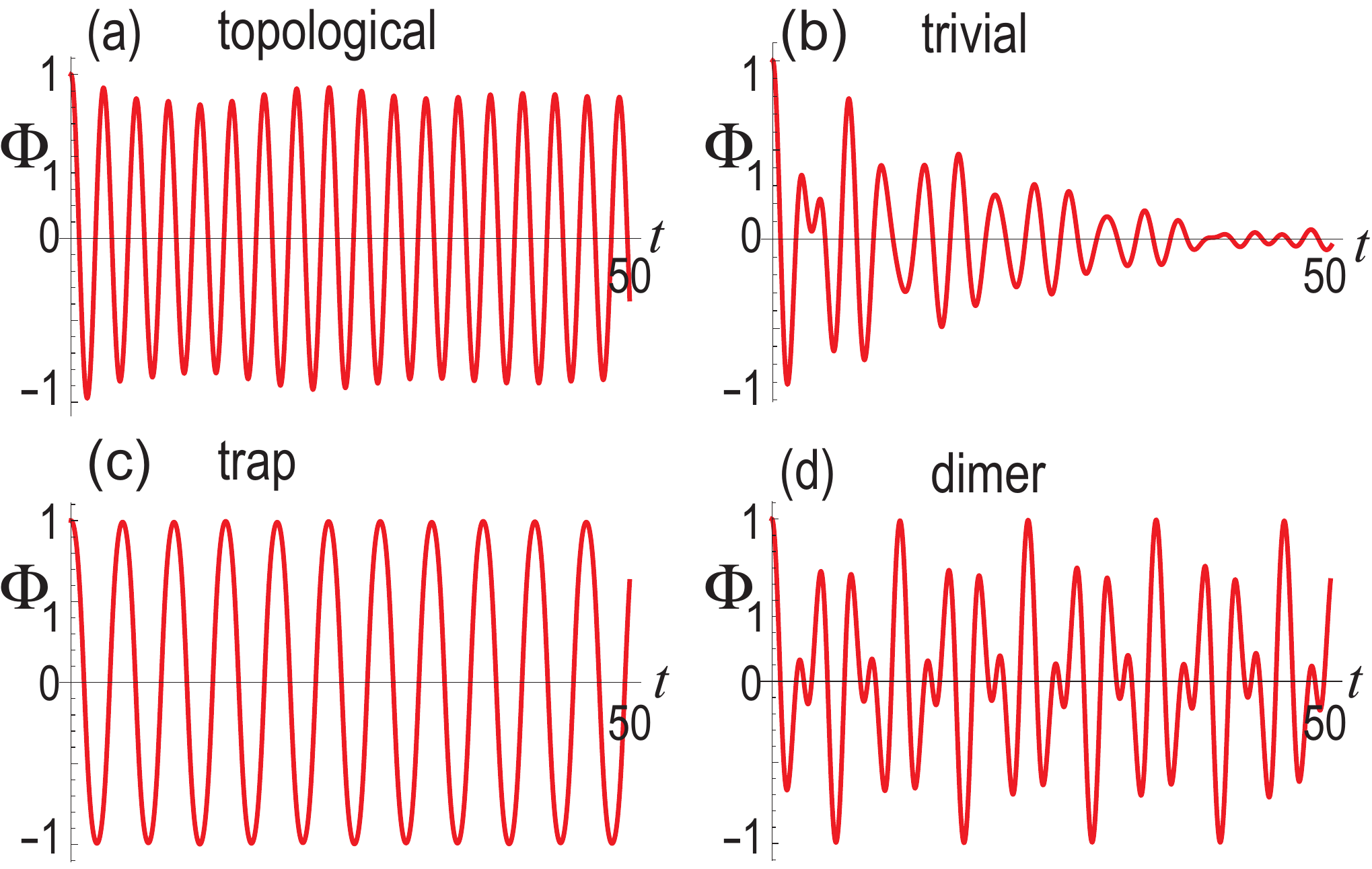}}
\caption{Time evolution of $\Phi _{1}$ in units of $\left( \protect\xi \Phi
_{0}/2\right) $. (a) A topological phase with $\protect\lambda =-0.5$ and 
$\protect\xi =0.25 $ (b) a trivial phase with $\protect\lambda =0.5$ and 
$\protect\xi =0.25$, (c) trap phase with $\protect\lambda =-0.8$ and $\protect\xi =0.8$ 
and (d) the dimer phase with $\protect\lambda =0.9$ and $\protect\xi =0.5$. 
We have set and $I_{\text{c}}=10$ and used a chain with the
length $100$ and set $T=50$.}
\label{FigEvolve}
\end{figure}

\textbf{Quench dynamics and phase diagram:} We investigate the quench
dynamics by taking the initial condition such as%
\begin{equation}
\Phi _{n}=\frac{1}{2}\xi \Phi _{0}\delta _{n,1},\quad \dot{\Phi}_{n}=0,
\label{IniCon}
\end{equation}%
with $0\leq \xi \leq 1$. Namely, we give the initial flux with the strength 
$\frac{1}{2}\xi \Phi _{0}$ only to the left-end CPB. Then, allowing it to
move with the zero initial velocity, we study how the motion propagates
along the array. We see later that the strength $\xi $\ controls the
nonlinearity of the system.

We have examined the quench dynamics by taking various values of $\lambda $
and $\xi $. We have found numerically that there are four types of
propagations, whose typical structures are given in Fig.\ref{FigDynamics}.
The time evolution at the edge, $\Phi _{1}\left( t\right) $, is shown in Fig.\ref{FigEvolve} for these four types of propagations.

In Fig.\ref{FigDynamics}(a), there are standing waves mainly at the left-end
CPB and weakly at a few adjacent odd-number CPBs. Furthermore, there are
propagating waves into other CPBs. In Fig.\ref{FigDynamics}(b), there are
only propagating waves into other CPBs. In Fig.\ref{FigDynamics}(c), the
standing wave is trapped strictly at the left-end CPB. In Fig.\ref{FigDynamics}(d), 
the coupled standing waves are trapped to double CPBs at
the left-end.

We next search for the phase diagram in the $\lambda $-$\xi $ plane. We
define the phase indicator by the saturated amplitude as%
\begin{equation}
\Psi \equiv \max_{T_{1}<t<T_{2}}\frac{\left\vert \Phi _{1}\left( t\right)
\right\vert }{\xi \Phi _{0}/2},  \label{SatuAm}
\end{equation}%
where $T_{1}$ and $T_{2}$ are taken much larger than the period and $%
T_{2}-T_{1}$ is of the order of the period of the modulation.

\begin{figure}[t]
\centerline{\includegraphics[width=0.48\textwidth]{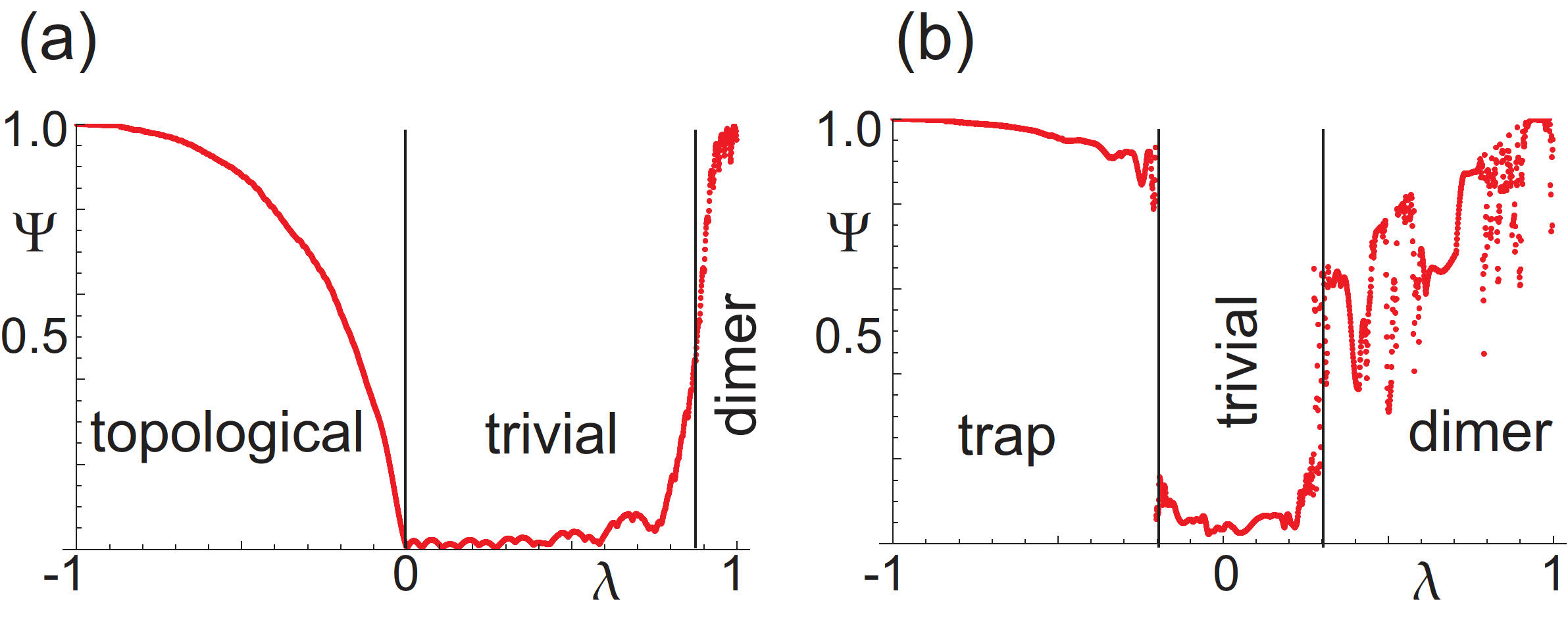}}
\caption{Phase indicator $\Psi $ as a function of dimerization $\protect\lambda $. 
(a) $\protect\xi =0.25$ and (b) $\protect\xi =0.8$. 
We have set $I_{\text{c}}=10$ and used a chain with the length $100$. 
We have set $T_{1}=45$ and $T_{2}=50$.}
\label{FigDistribute}
\end{figure}

\begin{figure}[t]
\centerline{\includegraphics[width=0.48\textwidth]{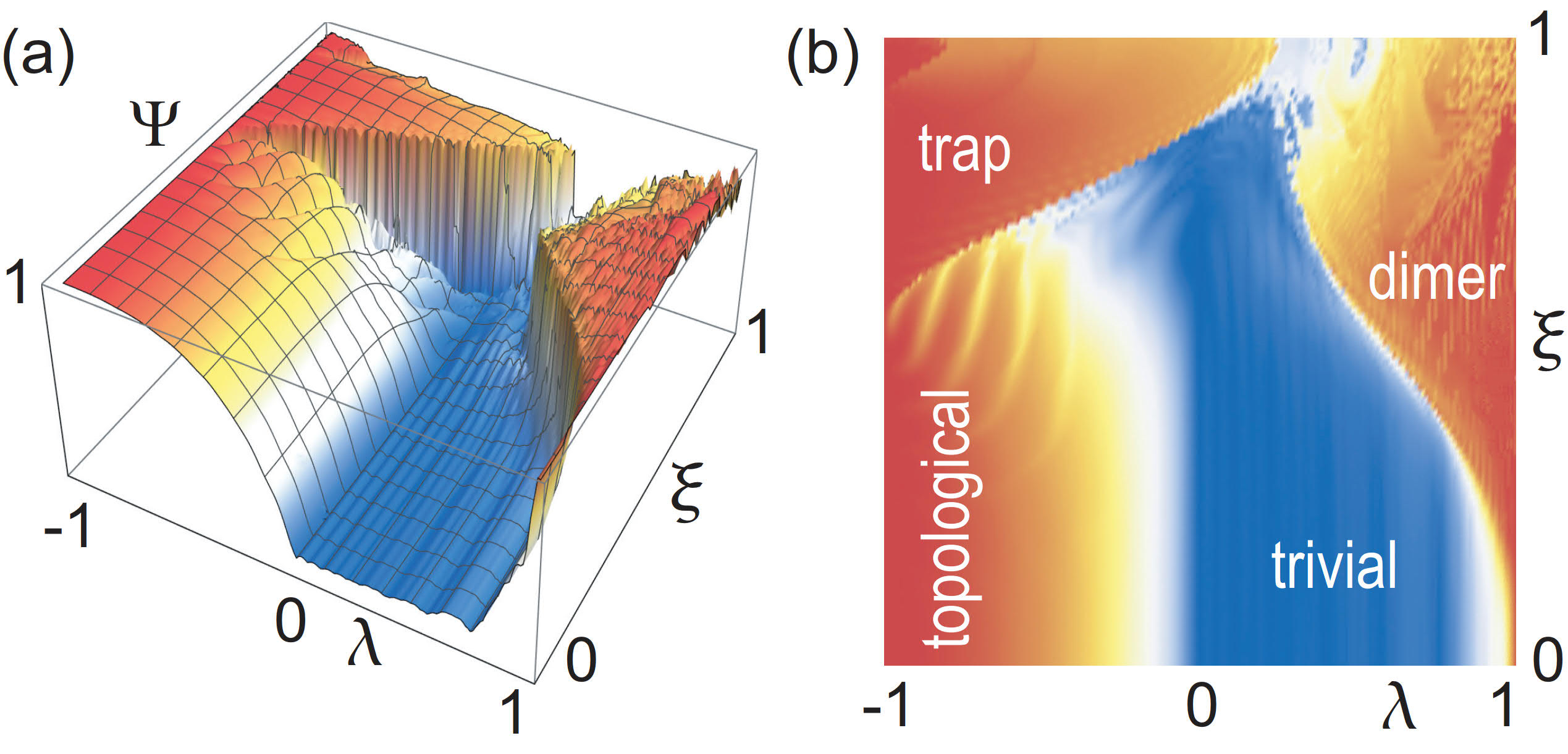}}
\caption{Phase indicator $\Psi $ in the $\protect\lambda $-$\protect\xi $
plane. (a) Bird's eye's view and (b) Top view. We have set $I_{\text{c}}=10$
and used a chain with the length $100$. We have set $T_{1}=45$ and $T_{2}=50$.}
\label{FigPhase}
\end{figure}

First, we show it as a function of $\lambda $ in Fig.\ref{FigDistribute}(a).
It exhibits a sharp transition at $\lambda =0$, which is finite for $\lambda
<0$ and almost zero for $\lambda >0$ in the weak nonlinear regime $\xi
\lesssim 1/2$. It must be a phase transition separating the two phases with 
$\lambda <0$ and $\lambda >0$. We also show $\Psi $ in the $\lambda $-$\xi $
plane in Fig.\ref{FigPhase}. Later we argue that they are the topological
and trivial phases. We find that the phase boundary remains as it is for $\xi \lesssim 1/2$.

There are two more phases with the phase transition points identified by
gaps in the phase indicator in Fig.\ref{FigDistribute}(b). One is present
around the upper-left corner of Fig.\ref{FigPhase}(b), which we call the
trap phase because the wave is trapped to the left-end CPB by the nonlinear
effect as in Fig.\ref{FigDynamics}(c). The other is present along the right
side, which we call the dimer phase since its origin is the dimer mode as in
Fig.\ref{FigDynamics}(d). We later ague that the former is well approximated
by a single CPB and the latter by double CPBs.

These four phases are distinguishable by the time evolution of propagating
modes. There is a stable oscillation although the amplitude is smaller than
1 in the topological phase as shown in Fig.\ref{FigEvolve}(a). The amplitude
rapidly decreases in the trivial phase as shown in Fig.\ref{FigEvolve}(b).
The oscillation with the amplitude 1 is observed in the trap phase as shown
in Fig.\ref{FigEvolve}(c). There is a beating oscillation in the dimer phase
as shown in Fig.\ref{FigEvolve}(d).

\textbf{Topological argument:} In order to clarify the topological property
of the dynamics, we study the hopping matrix $M_{nm}$\ in Eq.(\ref{MEq}). We
split the hopping matrix between the diagonal and the off-diagonal terms as%
\begin{equation}
M_{nm}=\left( C+2C^{\prime }\right) \delta _{nm}-\overline{M}_{nm},
\end{equation}%
where%
\begin{eqnarray}
\overline{M}_{nm} &=&[C_{A}^{\prime }\delta _{m,n+1}+C_{B}^{\prime }\delta
_{n,m+1}]\quad \text{for odd }n,  \notag \\
\overline{M}_{nm} &=&[C_{B}^{\prime }\delta _{m,n+1}+C_{A}^{\prime }\delta
_{n,m+1}]\quad \text{for even }n.
\end{eqnarray}%
Here, $\overline{M}_{nm}$ describes the SSH\ model, which is typical
topological model.

In the momentum space, it reads%
\begin{equation}
\overline{M}\left( k\right) =\left( 
\begin{array}{cc}
0 & q\left( k\right) \\ 
q^{\ast }\left( k\right) & 0%
\end{array}%
\right)  \label{Mk0}
\end{equation}%
with $q\left( k\right) =C_{A}^{\prime }+C_{B}^{\prime }e^{-ik}$. The
topological number is the Zak phase defined by%
\begin{equation}
\Gamma =\frac{1}{2\pi }\int_{0}^{2\pi }A\left( k\right) dk,
\label{ChiralIndex}
\end{equation}%
where $A\left( k\right) =-i\left\langle \psi (k)\right\vert \partial
_{k}\left\vert \psi (k)\right\rangle $ is the Berry connection with $\psi
(k) $ the eigenfunction of $M_{0}\left( k\right) $. It is well known that
the system is topological ($\Gamma =1$) for $C_{B}^{\prime }>C_{A}^{\prime }$
and trivial ($\Gamma =0$) for $C_{B}^{\prime }<C_{A}^{\prime }$ in the SSH
model.

\textbf{Linearized theory:} Although the hopping matrix $M_{nm}$ has a
topological structure, it is not clear how the topological property
manifests itself in the CPB array because it is a part of the kinetic term.
We now argue that the phase diagram in Fig.\ref{FigPhase} is just the
topological phase diagram.

We have found in Fig.\ref{FigPhase} that the phase boundary at $\lambda =0$
is almost identical in a wide region for $\xi <1/2$. We construct a
linearized theory near $\xi =0$, where the equations of motions read%
\begin{equation}
\sum_{m}M_{nm}\ddot{\Phi}_{n}+2\pi I_{\text{c}}\Phi _{n}/\Phi _{0}=0.
\label{EqB}
\end{equation}%
We diagonalize the hopping matrix as%
\begin{equation}
M_{nm}\Phi _{m}^{\left( p\right) }=E_{p}\Phi _{m}^{\left( p\right) },
\label{EqA}
\end{equation}%
where $p$ labels the eigenvalue. In the basis of the eigenfunction, the
equations of motion is given by%
\begin{equation}
E_{p}\ddot{\Phi}_{m}^{\left( p\right) }=-2\pi I_{\text{c}}\Phi _{m}^{\left(
p\right) }/\Phi _{0},
\end{equation}%
whose solutions are%
\begin{equation}
\Phi _{m}^{\left( p\right) }\left( t\right) =\exp \left[ \pm i\Omega _{p}t\right] \Phi _{m}^{\left( p\right) }\left( 0\right) ,
\end{equation}%
with the characteristic frequency $\Omega _{p}\equiv 2\pi I_{\text{c}}/\Phi
_{0}E_{p}$. This is also the solution of Eq.(\ref{EqB}).

The eigenfunction $\Phi _{m}^{\left( p\right) }$ is that of the SSH model 
$\overline{M}_{nm}$ as well. In the topological phase of the SSH model, there
are zero-energy edge states $\Phi _{m}^{\left( 0\right) }$ localized at the
edges ($\overline{E}_{0}=0$). They describe also the localized edge states
with the energy $E_{0}=C+2C^{\prime }$ in Eq.(\ref{EqA}). They remain at the
edge after the time evolution. On the other hand, there are no such
localized edge states in the trivial phase, and hence, there are no modes
remaining at the edge. These phenomena are the clear distinction between the
topological and the trivial phases in the present Cooper-box array. We have
numerically confirmed that they are valid even in the nonlinear regime.

\textbf{Topological and trap phases:} In Fig.\ref{FigDynamics}(a) and (c),
the dynamics is induced mainly at the left edge. Time evolution of $\Phi _{1}
$ is shown in Fig.\ref{FigEvolve}(a) and (c). In order to obtain analytic
understanding, we consider the limit $\lambda \rightarrow -1$, where the
left-end CPB is detached from the rest since $C_{A}^{\prime }=0$.

\begin{figure}[t]
\centerline{\includegraphics[width=0.48\textwidth]{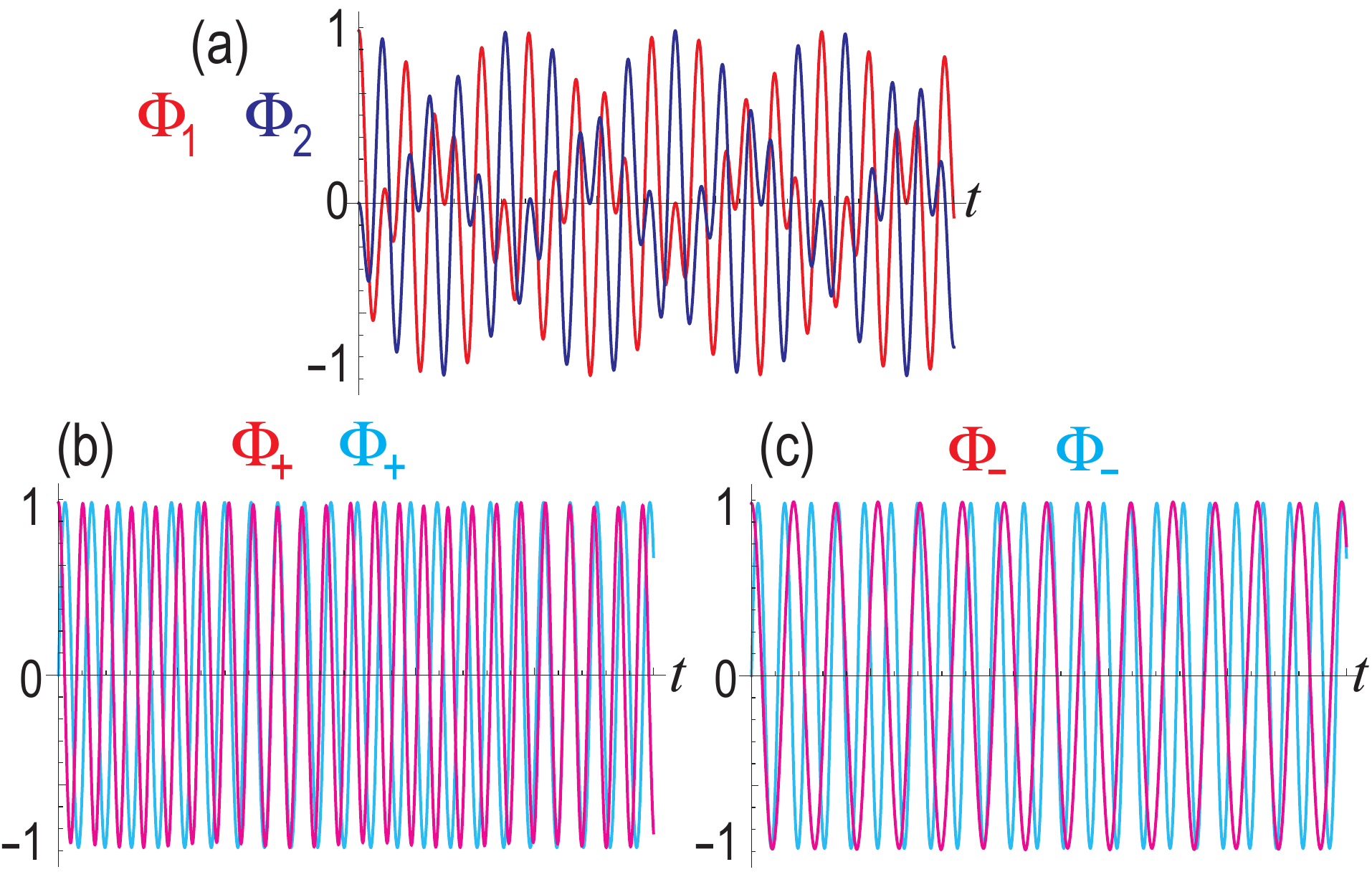}}
\caption{The dynamics of double Cooper-pair boxes. We have set $\protect\xi =0.1$. 
(a) $\Phi _{1}$ in red and $\Phi _{2}$ in blue. (b) $\Phi _{+}$ and
(c) $\Phi _{-}$. (b) and (c) The numerical results are show in magenta and
the analytic solutions (\protect\ref{Jacobi1}) and (\protect\ref{Jacobi2})
are shown in cyan. We have set $\protect\xi =0.25$ and $I_{\text{c}}=10$.}
\label{FigDimer}
\end{figure}

The equation of motion for the left-end CPB in Fig.\ref{FigCPBox}(c) is
given by%
\begin{equation}
\ddot{\Phi}_{1}=-(I_{\text{c}}/C)\sin 2\pi (\Phi _{1}/\Phi _{0}).
\label{StrongDSG}
\end{equation}%
The exact solution is given by%
\begin{equation}
\Phi _{1}(t)=2\Phi _{0}\sin ^{-1}[\alpha \sin \text{am}(\alpha ,\omega
t+K\left( k\right) )],  \label{PenduMotionA}
\end{equation}%
where am is the Jacobi amplitude function, $K\left( k\right) $ is the Jacobi
perfect elliptic integral of the first kind, $\omega \equiv \sqrt{I_{\text{c}}/C}$ 
and $\alpha $ is determined as $\alpha =\sin [\Phi \left( 0\right) /2]$
in terms of the initial condition $\Phi (0)$. The period is given by $T=4\omega ^{-1}K\left( k\right) $. 
This solution well describes the wave
function of the topological edge state numerically obtained.

The solution (\ref{PenduMotionA}) is valid at $\lambda =-1$ irrelevant of
the value of $\xi $. Hence, it describes the topological edge state in the
topological phase as well as the edge state in the trap phase. Indeed, we
have numerically checked that this single CPB wavefunction well describes
the edge state for the topological phase and the trap phase in Fig.\ref{FigEvolve}(a) and (c). 
Actually, there is no transition between the
topological and trap phases at $\lambda =-1$. However, numerical solutions
show that there is a sharp transition between the topological and trap
phases for $\lambda >-1$. The topological phase is smoothly connected with
that of the linear theory. However, the trap phase is not connected with any
phase in the linear theory. Especially, we cannot assign a topological
number for the trap phase because it is an isolated oscillation mode.

\textbf{Dimer phase:} In Fig.\ref{FigDynamics}(d), the dynamics is induced
solely at the left two edges. Time evolution of $\Phi _{1}$ is shown in Fig.\ref{FigEvolve}(d). 
In order to obtain analytic understanding, we consider
the limit $\lambda \rightarrow 1$, where the left-end double CPBs are
detached from the rest since $C_{B}^{\prime }=0$.

We analyze the dynamics of a pair of CPBs connected via a capacitor\cite{Krantz} 
as shown in Fig.\ref{FigCPBox}(b). The equations of motions read%
\begin{align}
\left( C+C^{\prime }\right) \ddot{\Phi}_{1}-C^{\prime }\ddot{\Phi}_{2}& =-I_{%
\text{c}}\sin 2\pi (\Phi _{1}/\Phi _{0}), \\
\left( C+C^{\prime }\right) \ddot{\Phi}_{2}-C^{\prime }\ddot{\Phi}_{1}& =-I_{%
\text{c}}\sin 2\pi (\Phi _{2}/\Phi _{0}).
\end{align}%
We numerical solve these equations with the initial condition (\ref{IniCon}%
), whose result is shown in Fig.\ref{FigDimer}(a). The beating behavior is
observed. We plot $\Phi _{+}\equiv \Phi _{1}+\Phi _{2}$ and $\Phi _{-}\equiv
\Phi _{1}-\Phi _{2}$ in Fig.\ref{FigDimer}(b) and (c). The equations of
motion reads 
\begin{align}
\ddot{\Phi}_{+}& =-(I_{\text{c}}/C)\sin \pi (\Phi _{+}/\Phi _{0})\cos \pi
(\Phi _{-}/\Phi _{0}), \\
\ddot{\Phi}_{-}& =-(I_{\text{c}}/\left( C+2C^{\prime }\right) )\cos \pi
(\Phi _{+}/\Phi _{0})\sin \pi (\Phi _{-}/\Phi _{0}),
\end{align}%
whose approximate solutions are%
\begin{align}
\Phi _{+}\left( t\right) & =2\Phi _{0}\sin ^{-1}[\alpha \sin \text{am}%
(\alpha ,\omega _{+}t+K\left( k\right) )],  \label{Jacobi1} \\
\Phi _{-}\left( t\right) & =2\Phi _{0}\sin ^{-1}[\alpha \sin \text{am}%
(\alpha ,\omega _{-}t+K\left( k\right) )],  \label{Jacobi2}
\end{align}%
with $\omega _{+}\equiv \sqrt{I_{\text{c}}/C}$, $\omega _{-}\equiv \sqrt{I_{\text{c}}/(C+2C^{\prime })}$ 
and $\alpha \equiv \sin [\Phi \left( 0\right)
/2]$. The period of $\Phi _{+}\left( t\right) $ is $T=4K\left( k\right)
/\omega _{+}$ and that of $\Phi _{-}\left( t\right) $ is $T=4K\left(
k\right) /\omega _{-}$. In numerical simulations, there are fluctuation of
the period although the overall period is identical to the analytic solution.

We have found a dynamical topological phase transition in a CPB array, which
is absent in static system. It is highly contrasted to standard models,
where the topological phase is present in static system. It is possible to realize the present system by using existing superconducting qubit systems.

The author is very much grateful to N. Nagaosa for helpful discussions on
the subject. This work is supported by CREST, JST (JPMJCR20T2).

\end{document}